\documentstyle[preprint,epsf,aps]{revtex}
\begin{document}
\draft
\title
{Model for the low-temperature magnetic phases 
          observed in doped YBa$_2$Cu$_3$O$_{6+{\rm x}}$}
\author{Niels Hessel  Andersen}
\address{Department of Solid  State  Physics,  Ris{\o} National 
Laboratory,   DK-4000   Roskilde,    Denmark}
\author{Gennadi Uimin}
\address{Institut f\"ur Theoretische Physik, 
Universit\"at zu K\"oln, Z\"ulpicher Str.77, 
D-50937 K\"oln, Germany\\
and\\
Landau Institute for Theoretical Physics, Chernogolovka, 142432
Moscow District,  Russia}
\maketitle

\begin{abstract}
A classical statistical model for the antiferromagnetic (AFM) ordering 
of the Cu-spins in the CuO$_2$  
planes of reduced YBa$_2$Cu$_3$O$_{6+{\rm x}}$  type materials is presented. 
The magnetic phases considered are the experimentally  observed  
high-temperature  AFI  phase
with ordering vector ${\bf Q}_{\rm I}=(\frac{1}{2} \frac{1}{2} 0)$, and
the low-temperature phases: AFII  with
${\bf Q}_{\rm II}=(\frac{1}{2} \frac{1}{2} \frac{1}{2})$ and 
intermediate  TA   ({\it  Turn  Angle}) phases TAI, 
TAII and TAIII  with components  of  both  ordering vectors.
It is shown that the AFII and TA phases
result from an effective ferromagnetic (FM) type coupling mediated 
by free spins in the CuO$_x$ basal plane.
Good agreement  with experimental data is  obtained for realistic
model parameters.
\\

\end{abstract}

It is  well-established  that  the  CuO$_2$  layers  in  the
high-temperature     superconductor     materials, like
La$_2$CuO$_4$ and YBa$_2$Cu$_3$O$_{6+{\rm x}}$  (YBCO),  form  an  
{\it antiferrromagetic} (AFM)  state
when   they    are insufficiently hole  doped  for
superconductivity \cite{Tranquada88}.  
The  nearly  2$d$ quantum  state  of   these   magnetic   phases   makes  them
intrinsically interesting, but a major  motive  for  studying
their properties is clearly the unsettled question,  whether
the magnetic fluctuations contribute to establish the
superconducting state or compete with and eventually destroy it. 

          Despite the many detailed studies of the static and  dynamic
          properties of the  magnetic  phases  in  YBCO type materials,
          the origin of the low-temperature  AFII  magnetic
          phase has not been satisfactorily understood. The AFII phase
          has been found in the same range of  oxygen  stoichiometries
          $(x < 0.35)$ as the high temperature AFI phase in  apparently
          pure YBCO \cite{Kadowaki88,Shamoto93} and the homologous rare earth 
          (RE) substituted REBCO \cite{Moudden88,Li90}. However, 
           recent  experimental  studies  have shown that it is {\it not}
            present  in  high-purity  single crystals of YBCO 
            \cite{Casalta94,Brecht95} and NdBCO \cite{Brecht97},
             whereas it is found in  Al \cite{Casalta94,Brecht95},
              Co \cite{Miceli89} and Fe \cite{Mirebeau94} doped YBCO,
           and in YBCO with La doped onto the Ba site \cite{Schmidt97}.
            The early  observations  of  the  AFII  phase  in
          nominally pure YBCO single crystals \cite{Kadowaki88,Shamoto93} 
          may therefore result from corrosion of the alumina crucibles used 
          for the crystal growth.
          
          Experimentally  it  has  been  shown   that   the  Cu magnetic
          structures in  YBCO    materials  are characterized by 
a simple AFM ordering of the Cu spins within the CuO$_2$ double layers 
as shown in Fig.~1a.
The Cu spins are lying  in  the  plane  but  the  in-plane  easy
          direction has  not  been  established  yet. 
Well below  the  N\`eel temperature,
           ($T_N\!=\!410$ K for $x \approx 0$), the Cu$^{2+}$ spins in the
            CuO$_2$  double layers form a rigid collinear 2$d$ structure.
In the AFI phase  adjacent  double  layers ({\it i.e. next
       nearest} CuO$_2$ planes) are AFM coupled via the CuO$_x$ basal plane.
Thus compared  to  the  chemical unit  cell, the magnetic  unit
          cell of the AFI structure is doubled along the $a$  and  $b$
          axes, but not along the $c$ axis giving an  ordering vector 
    ${\bf  Q}_{\small{\rm I}}=(\frac{1}{2}  \frac{1}{2}  0)$.  In   the   AFII
          structure the spins on next nearest CuO$_2$  planes
          are aligned by an effective  {\it ferromagnetic} (FM) coupling  via  the
          CuO$_x$ basal plane  with  a  magnetic  unit  cell doubled along the
$c$ axis, and hence 
${\bf Q}_{\small{\rm II}}=(\frac{1}{2} \frac{1}{2} \frac{1}{2})$ 
\cite{Kadowaki88,Shamoto93,Casalta94,Brecht95}.
          In an intermediate phase, which we shall denote the TA 
({\it  Turn Angle})  phase,  components  of   both vectors,
${\bf Q}_{\small{\rm I}}$ and ${\bf Q}_{\small{\rm II}}$, are observed  
experimentally.
          It has been debated whether the TA phase is a mixed AFI--AFII
          phase or results from a  continuous  transformation between the
          two spin configurations.
          In this letter we establish a  classical  statistical
          model that explains the origin of  the  interaction  leading
          to the AFII phase at low temperatures,  and  the  transition
          to the AFI phase via the TA phase.  
For pure YBCO  there is  essentially
          no free magnetic moment on the Cu sites  in  the  CuO$_x$
          basal plane even for the oxygen content close  to  the
           transition into the superconducting state, $(x\approx 0.3)$,
            where significant amount of magnetic  Cu$^{2+}$
          ions should be present.
This may  be  explained  by
          the formation of randomly distributed singlet spin-pairs  of
          Cu$^{2+}$--O$^{2-}$--Cu$^{2+}$   fragments,    which    like
          isolated Cu$^+$ ions are  {\it  non-magnetic},  and  onwards
          shall    be    referred    to    as    such.    Not only shortest,
           but also longer chain fragments, like Cu$_3$O$_2$,
            should be considered as non-magnetic because the holes may
            redistribute inside the chains and leave the spin state 
of such a "molecule" as a singlet.    
It is argued that free Cu spins may be formed by direct substitution of
{\it magnetic} ions like Co \cite{Miceli89} and Fe \cite{Mirebeau94} 
for {\it non-magnetic} Cu, or indirectly
when Al$^{3+}$ substitutes for Cu \cite{Casalta94,Brecht95} or RE$^{3+}$ for 
Ba$^{2+}$ \cite{Brecht97,Schmidt97}. In the latter case, which may occur for 
light RE, it is likely that additional oxygen is introduced into the basal 
plane to assure charge neutrality and gives rise to additional oxidation of 
Cu$^+$ to Cu$^{2+}$. With RE$^{3+}$ substitution for Ba$^{2+}$ one free 
Cu$^{2+}$ spin is expected to result, whereas Al$^{3+}$ may give rise to two
Cu$^{2+}$ spins, {\it e.g.} as shown in Fig.~1a. These two
Cu$^{2+}$ spins are essentially free because they are linked via {\it 
non-magnetic} Al or the O-O orbital overlap. 

The  free  spins  may  be  polarized   and   establish   an   effective
          FM coupling between the rigid spin configurations
          of two adjacent CuO$_2$ double layers at  low  temperatures.
          For sufficiently high  concentrations  of  free  spins  the 
          FM coupling  will  dominate  and  form  the  AFII
          phase, but at higher temperatures it gradually  becomes
          ineffective and the  ordering  will  be   controlled  by the
          AFM couplings   mediated   by    the    {\it
          non-magnetic} Cu ions  ({\it  cf.}  Fig.~1a).
          Thereby the AFII structure is transformed into the TA  phase
          and further into the AFI structure via two continuous phase
          transitions. For low concentrations of free spins  the  AFII
          phase does not form and the TA phase is  the  stable  ground
          state.
           Including a fourfold easy-axis anisotropy in addition
           to a well-established in-plane anisotropy  for Cu
          spins in the bilayers first order  phase transitions to various
           TA phases, TAI, TAII  and TAII may result. 

          The total Hamiltonian for the spin system can be represented 
          via the sum ${\cal H}_{\rm tot} = \sum^5_{k=0} {\cal H}_k$, 
and the relevant interaction parameters are defined in Fig.~1.
${\cal H}_0$ is the intra-plane Heisenberg  spin-exchange
and  ${\cal  H}_1$  is the Heisenberg  exchange  between  the Cu spins in 
the nearest CuO$_2$ planes. ${\cal H}_2$ is the easy-plane anisotropy and 
${\cal H}_3$ reflects the anisotropy for spin rotation in the easy-plane.
   The indirect Heisenberg exchange interactions 
between  Cu  spins in the next nearest CuO$_2$ planes  via {\it  non-magnetic}
Cu ions, and via free  spins in the  basal  CuO$_x$
plane, are given by Hamiltonians ${\cal H}_4$  and
${\cal H}_5$, respectively. We do not include 
the  interactions between spins in  the  CuO$_x$  basal  plane,  
which may be estimated to have little influence for low 
concentrations of free spins.

In principle the model is applicable for the above mentioned dilute doping 
mechanisms and their combinations. However, in the following we shall focus on 
the properties of Al-doped YBCO. We shall consider any basal plane as 
consisting of free Cu$^{2+}$ spins, of {\it  non-magnetic} Cu ions, and of 
Al-occupied sites with relative concentration, $x$, $y$ and $\delta$, that 
satisfy the equality $x + y + \delta = 1$. {\it Non-magnetic}  Cu and Al 
ions are supposed to be randomly distributed, while free Cu$^{2+}$ spins  
correlate spatially with Al, {\it e.g.} as shown in 
Fig.~1a. The Cu spins in the $j$-th  bilayer which form a rigid 2$d$
collinear AFM structure are characterized by a classical spin ${\bf  S}_j$, 
that is the $j$-th layer order parameter.
The justification for  this  assumption  is that
the values  of  the  intra-plane  and  inter-plane  coupling
          constants, ${\cal J}_0 \approx 170$ meV in ${\cal  H}_0$  and
          ${\cal J}_1 \approx  10^{-1}{\cal J}_0 $ in  ${\cal  H}_1$, 
significantly exceed all other contributions to ${\cal H}_{\rm tot}$ 
\cite{Rossat-Mignod93,Hayden96} ({\it cf.} 
the experimentally observed  temperatures  of
the AFII and TA ordering: 
$T_{\small{\rm AFII}} \leq T_{\small{\rm TA}} \leq 20$ K in YBCO). 
To be specific we ascribe ${\bf S}_{j}$ to the lattice site 
${\bbox\rho}=(0,0)$ of the bottom plane ($b$) of the $j$-th double layer.
Then, evidently, the spin-field in the bottom plane is 
${\bf S}_{j}^{(b)}({\bbox\rho})={\bf S}_{j}
\exp (\imath {\bf Q}{\bbox\rho})$ and in the top ($t$) plane
${\bf S}_{j}^{(t)}({\bbox\rho})=-{\bf S}_{j}^{(b)}({\bbox\rho})$.
Each spin of the set $\{{\bf  S}_j\}$ governed by 
the main contribution  ${\cal H}_{\rm tot}^{(0)}={\cal H}_0+{\cal  H}_1$ 
is still freely oriented, but this degeneracy is removed by smaller
interactions, that regulate the relative orientations of adjacent double 
layer spins and their orientations with respect to the easy axes. 
Because the energy scale of ${\cal H}_2$, ${\cal H}_3$ and ${\cal H}_4$
is much below that of the  temperature it is allowed to replace them by 
their "mean"-field expressions in the free energy. Thus ${\cal H}_2$ restricts
the spins to lie in the basal plane, and ${\cal H}_3$ leads to the  fourfold 
easy-axis anisotropy given by:
\begin{equation}
{\cal F}_3  = -{\cal D}_4 \sum_{\bf r} \cos 4\phi_{\bf r} 
\rightarrow  
-2N_{\rho}{\cal D}_4\sum_j\cos 4\phi_j
\label{eq:F_3} 
\end{equation}
where $\phi_j$ is the angle between ${\bf  S}_j$
and one of the fourfold easy axes, which, in lack of information we take to be 
the $x$ (or equivalently the $y$) axis.
$N_{\rho}$ is the total number of Cu sites in a CuO$_2$ plane. Further, 
\begin{equation}
{\cal F}_4= {\cal  J}_2  \sum_{j,{\bbox\rho}'} 
{\bf  S}_{j}^{(t)}({\bbox\rho}')
          {\bf S}_{j+1}^{(b)}({\bbox\rho}') 
          \rightarrow - y N_{\rho}{\cal  J}_2 \sum_{j}{\bf  S}_j{\bf  S}_{j+1}
\label{eq:F_4} 
\end{equation}
where the sum runs over the double layers, $j$, and over the $N_\rho y$ 
in-plane sites ${\bbox\rho}'$ which are  bridged  by {\it non-magnetic}
Cu  ions  in   the   basal   plane. 
          
The Hamiltonian ${\cal H}_5$ realizes coupling via {\it free} Cu spins:
\begin{equation}
          {\cal H}_5 =  {\cal J}_3 
\sum_{j,{\bbox\rho}'\!'}
          {\bbox\sigma}_{\!j+\frac{1}{2}}({\bbox\rho}'')
({\bf S}_{j}^{(t)}({\bbox\rho}'')+{\bf S}_{j+1}^{(b)}({\bbox\rho}''))
\label{eq:H_5}
\end{equation}
where now the summation is over $j$ and the $x N_\rho$ sites  ${\bbox\rho}''$ 
which are bridged by free Cu spins, ${\bbox\sigma}_{\!j+\frac{1}{2}}$. The 
corresponding free energy obtained from a classical statistical average 
over the angular degrees of freedom of ${\bbox\sigma}_{\!j+\frac{1}{2}}$ is:  
          \begin{equation}
{\cal F}_5 = - x N_{\rho} k_B T\ln 
\frac{\sinh(|{\cal J}_3|\,\sigma\,|{\bf S}_j -{\bf S}_{j+1}|/(k_B T))}
{|{\cal  J}_3|\, \sigma\, |{\bf S}_j- {\bf S}_{j+1}|/(k_B T)} \label{eq:F_5}
\end{equation}
which behaves as $- x N_{\rho}|{\cal J}_3|\, \sigma\,|{\bf  S}_j- {\bf S}_{j+1}|$
when $T\to 0$ and goes to zero for $T\to \infty$.
          Thus we arrive at the effective free energy, 
${\cal F}_{\rm eff}=\sum_{i=3}^5{\cal F}_i$, which takes the form of
an effective 1$d$  spin-system $\{{\bf S}_j\}$. 
Then our goal is to {min\-imize} 
${\cal F}_{\rm eff}$.
The polarization type of coupling leading to ${\cal F}_5$
tends to align the order parameters 
${\bf S}_j$  and ${\bf S}_{j+1}$ in antiparallel at low temperatures, 
whereas it becomes  ineffective
          at high temperatures, where  the parallel alignment
 of ${\bf S}_j$  and ${\bf S}_{j+1}$  will  prevail, due to ${\cal F}_4$
 if ${\cal J}_2\!>\!0$.  
At intermediate temperatures the competition between AFI (parallel) 
and AFII (antiparallel) alignments of the spins in adjacent 
bilayers  will lead to the TA phases. 
To show this we note, that the in-plane angles may be presented by
\begin{equation} 
\phi_j=\psi_0+(-1)^j\psi
\label{eq:spinconf} \end{equation}
where $\psi$ is the turn angle and $\psi_0=n\frac{\pi}4$. In $\psi_0$
{\it odd} $n$ (TAII phase) as well as {\it even} $n$ (TAI and TAIII phases) 
are necessary to establish all the minima of ${\cal F}_3$. Fig.~1b
illustrates the various possibilities, including the pure states, AFI and 
AFII, corresponding to $\psi$ equal to 0 and $\frac{\pi}2$, respectively. 
Note, that Eq. (\ref{eq:spinconf}) defines the spin rotation angles as 
strictly alternating: $\Delta\phi_j=\phi_{j+1}-\phi_j=
(-1)^{j\!-\!1}\cdot 2\psi$. For ${\cal D}_4=0$, this is not justified
because $\Delta\phi_j$ may take the form of degenerate configurations 
$R_j\cdot 2\psi$ with $R_j$ being a random set of 
$\pm 1$. As a result the spin configuration becomes 
disordered along $z$. A similar disordering  occurs in the TAII phase for 
$\psi =\frac{\pi}4$, which means $\Delta\phi_j=R_j\cdot\frac{\pi}{2}$ and 
is just the case shown in Fig.~1a. In principle, longer range interactions
 prevent the above mentioned  disordering to happen. $\psi =\frac{\pi}8$
  or $\frac{3\pi}8$
would also be candidates for disordering but because they appear at first
order transitions between TAII and  TAI or TAIII phases, this 
kind of a disordering is inaccessible.
Thus we arrive at the normalized free energy functionals 
($f^\pm = {\cal F}_{\rm eff}/(2 N_\rho | {\cal J}_3 | S \sigma)$):
\begin{equation}
          f^{\pm}(z)= y j z^2 \pm q(z^2-z^4) \mp\frac q8 - xt
          \ln\frac{\sinh(z/t)}{z/t}
\label{eq:function}
\end{equation}
                    where the normalized parameters are:
\begin{equation}
          j = \frac{{\cal J}_2 S}{|{\cal J}_3|  \sigma},  \;  \;  q  =
          \frac{8  {\cal  D}_4}{|{\cal  J}_3|  S  \sigma},  \;  \;  t  =
          \frac{k_B T}{2 |{\cal J}_3| S \sigma}, \; \; z =  \sin\psi
          \label{eq:norm} 
\end{equation}
          and $f^+$ and $f^-$ refer to the turn angles centered around the 
easy (TAI and TAIII phases) and hard (TAII phase) directions, respectively.

           For the concentration of free spin, $x$,  we  shall  assume that 
           each Al$^{3+}$ ion generates  two  free  Cu  spins, {\it i.e.}
           $x=2\delta$  ({\it cf.} Fig.~1a) and thereby:
           $y = 1 - \frac{3}{2} x$. 

         $f^\pm (z)$  have  been  minimized  with  respect  to $z$ as function 
of parameters, $t$ (reduced temperature) and    
$x$,  for  a few  values  of   the normalized
          interaction parameters $j$ and $q$. A realistic estimate for 
${\cal  J}_2$ is  $\sim 0.1$ K. Much less is known about 
          ${\cal J}_3$ and ${\cal D}_4$. We shall leave ${\cal D}_4$ as a 
          free parameter. ${\cal J}_3$ cannot be too small, because like in the 
          planes this spin interaction
is mediated via oxygen ions. However, hybridization via apical oxygens
is not as effective as hybridizations within the planes or chains because
the involved orbitals are ortogonal. Since the 
AFI - AFII transition temperature is of order a few Kelvin
we estimate ${\cal J}_3$ as 10 K and thereby: $j = 0.01$ and $T = 5 t$ K.

        Fig.~2 shows the corresponding phase  diagram
          for three values of  the  anisotropy  parameter  $q$.  Full
          lines represent continuous  transitions,  dotted  lines  are
first order transitions, and dots are tricritical points. The dashed lines are 
the {\it disorder line} at $z = \frac{\pi}4$. Although  the  anisotropy 
$q$ does  not change the phase diagram boundaries significantly, it
influences the  nature of the phase transitions even when $q\ll j$.
The AFI--TAI transition is always continuous, and if $q  = 0$, 
the AFI--AFII transition always
   takes place via two continuous transitions  through  the  intermediate 
 TA phase.  
For  low  concentrations  of  free  spins ($x\!<\!2jy$),
the pure AFII state is inaccessible even at $t\to 0$ (see Fig.~2). 
Increasing $j$ to 0.1 does  not  change
          the morphology of the phase diagram but for obvious  reasons
          higher concentrations of free spins  are  required  for  the
          characteristic features to occur.     
          One can subdivide  the 
order  parameter in the TA phases 
into AFI and AFII parts as $m_{\rm AFI}= \frac 12
|{\bf S}_j+{\bf S}_{j+1}| = S \sqrt{1 -z_g^2}$ and $m_{\rm AFII} =\frac 12
|{\bf S}_j-{\bf S}_{j+1}|= Sz_g$. 
          $z_g$ is the $z$ value at which the {\it global} minimun of the free
          energy is achieved from $f^\pm$ in 
Eqs.(\ref{eq:function}). 
The inserts in Fig.~2 show examples of the temperature variation of the 
{\it global} order parameter, $z_g$, as composed from the optimal $z^{\pm}$. 
Without anisotropy $(q=0)$ continuous transitions are observed, but for 
$q=0.001$ and $x=0.02$ a continuous transition from  AFI ($z_g=0$) to  TAI
($z_g=z^+$) is followed by the first order transition to TAII ($z_g=z^-$) 
and then to AFII ($z_g=1$).        
          
A  direct  comparison  with  experimental  data  requires  a
          detailed knowledge of the number of free spins generated for
          specific Al  doping  levels.  It  is  well-established  that
          the  sample  treatment  is  very  essential   for   the   Al
          configurations in  the  basal  plane,  and  quite  different
          concentrations of free spins may develop  for  the same Al 
          stoichiometry \cite{Brecht96}. However, comparing the model results
  displayed in Fig.~2  with experimental data 
          reported for  Al doped YBCO \cite{Casalta94,Brecht95} 
          we  find   agreement with the following observations:

(1) the sequental transitions AFI--TA--AFII occur below  20 K
({\it i.e.} below $t \simeq 4$); 
 
(2) the AFII  and  TA  phases
          are not present in nominally pure materials ({\it i.e.}  with
          $x=0$); 
 
(3) the transition into the AFII phase is  not  always
          complete  and  the  TA   phase   remains   stable   at   low
          temperatures and small $x$; 
 
(4) there is essentially  no  ordered  magnetic
          moments  on  the  Cu  sites  in  the  basal  plane; 
 
(5) the  intensities of magnetic Bragg peaks at
${\bf Q}_{\rm I}$ and ${\bf Q}_{\rm II}$ are in good agreement with the square 
of the order parameters $m_{\rm AFI}$ and  $m_{\rm AFII}$, ({\it cf.} insert 
to Fig. 2a). 

Since apparently no first order  transitions or jumps in the 
$m_{\small\rm AFI}$ 
and $m_{\small\rm AFII}$ order parameters are observed experimentally,  
the four-fold easy axis anisotropy is very small.

When large amount of free spins are introduced in the basal plane, as may be 
the case when YBCO is heavily doped with Co \cite{Miceli89} or Fe 
\cite{Mirebeau94}, they interact and it 
is likely that they form a 2$d$ AFM structure. In this case the AFII phase
is expected to be stable even at higher temperatures as has been
observed experimentally in Fe doped YBCO, where complete suppression of 
the AFI phase may result \cite{Mirebeau94}. 
  
           A classical treatment of  a  spin-$\frac{1}{2}$  system  may
          appear  inappropriate.  However,  it  should  be   recalled,
          that we are mainly giving a qualitative mechanism
          for the origin of the AFI , AFII and TA phases.  A   quantum
          treatment will be  pursued. 
           
           In conclusion we have shown by a classical statistical model
          that    the  AFII  and  TA  phases,  observed
          experimentally in Al, Co and Fe  doped  YBCO and in materials 
          with rare earth (RE) ions  on the Ba site,  result  from 
          polarization  of  free  spins  created in the basal plane of the
          structure. Thus, we propose that in nominally pure materials with
          no RE on the Ba site the AFII phase should not be present.
            Using  realistic values  for  the
          interaction parameters a semi-quantitative agreement with  the
          experimental observations is established. Based on the experimental
          fact that the magnetic transitions appear to be continuous and 
          long range order is etablished 
          along the $z$ axis we conclude that the in-plane 
          anisotropy parameter, ${\cal D}_4$ is finite but very small 
          $(<10^{-5} |{\cal J}_3|$, {\it cf.} Fig. 2b).
         More  model  work
          and experimental studies on carefully prepared materials   are
          needed and in progress for more detailed comparisons.

\begin{figure}
\epsfxsize=180mm
\epsffile{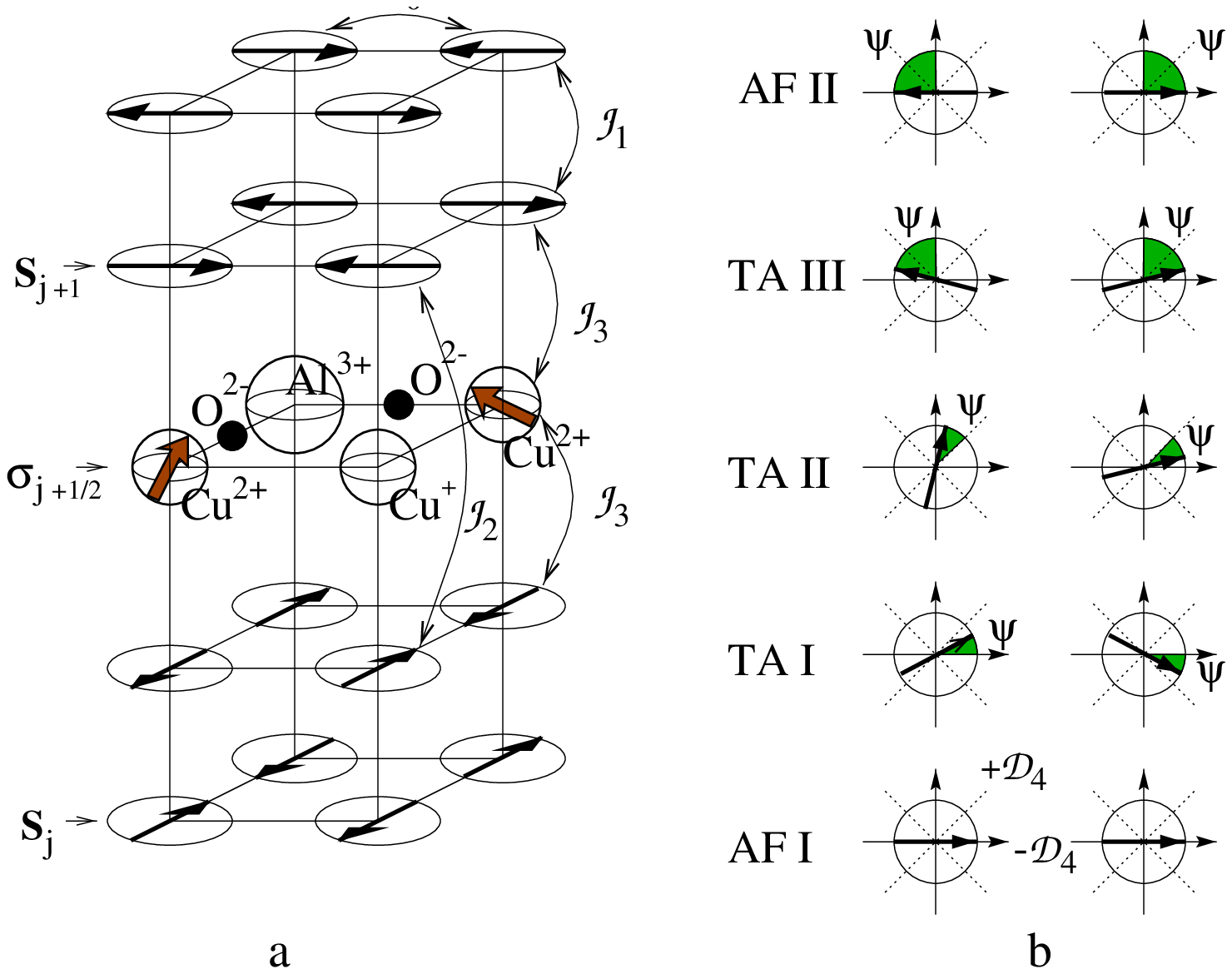}

\baselineskip=10pt
{{\small FIG.\ 1\quad Magnetic structures observed in YBCO. (a)  
Interactions ${\cal J}_0$ and ${\cal J}_1$ are responsible for the rigid 
double layer magnetic structures; {\bf S}$_j$ is the bilayer order parameter;
${\cal J}_2$ the coupling via non-magnetic Cu$^+$ that favors AFI structure, 
while ${\cal J}_3$ via the free Cu$^{2+}$ spins tends to AFII ordering.
The formation of Cu$^{2+}$ spins through Al substitution is also shown. 
The mutual spin arrangement in the bilayers is related to the TAII 
phase. (b) displays spin configurations and turn angles (shaded areas) 
$\psi$ in adjacent layers $j$ and $j+1$; magnetic phases 
AFI, AFII as well as various TA phases may result from the model studies. 
$\pm{\cal D}_4$ are related  to hard and easy axis spin anisotropies.}}
\end{figure}
\begin{figure}
\epsfxsize=128mm
\hspace{1.5cm}
\epsffile{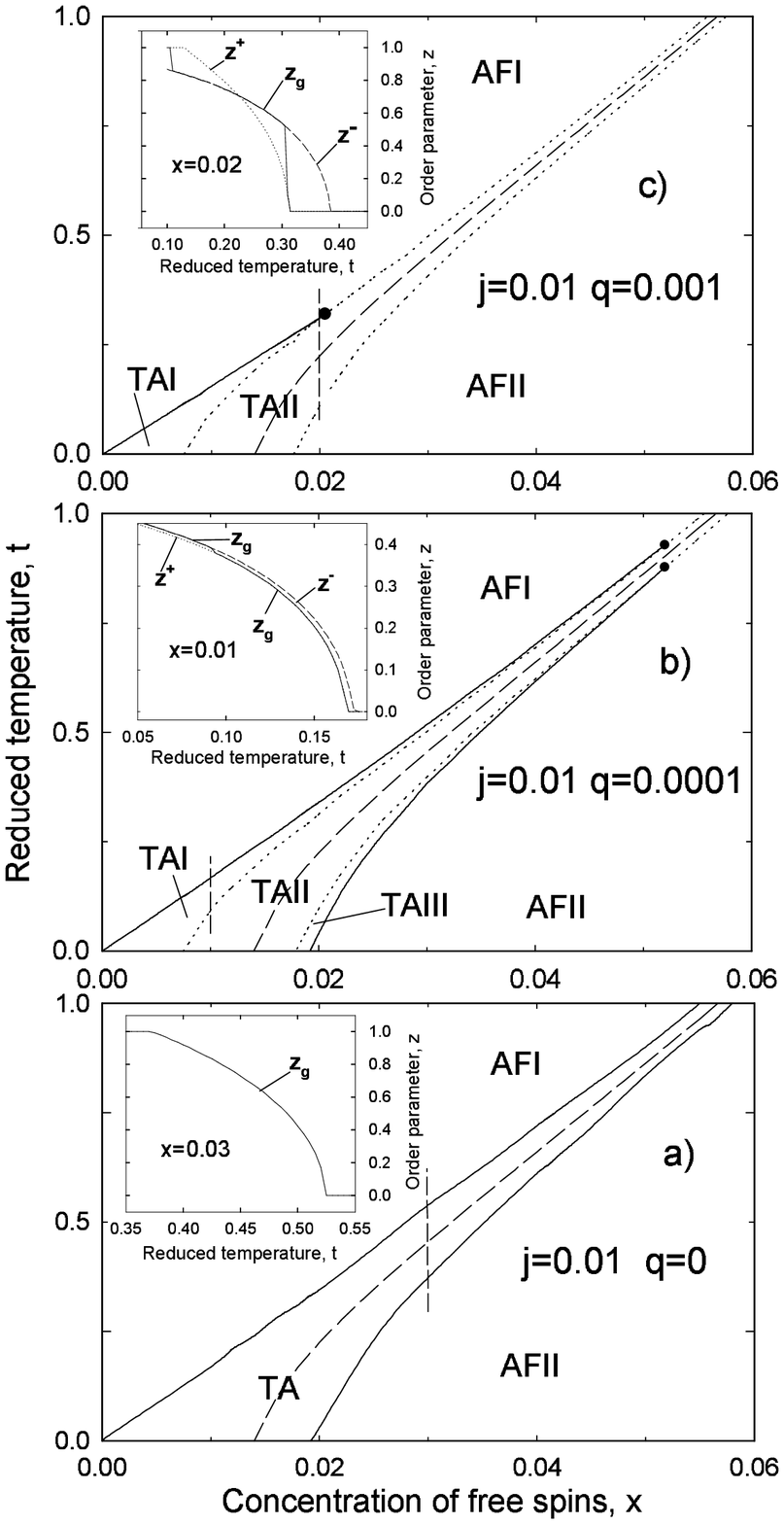}
\hfill

\baselineskip=10pt
{{\small FIG.\ 2\quad Magnetic phase diagrams for  Al  doped
      YBCO as function of the free spin concentration $x$ and the reduced
          temperature, $t$, for the  normalized  interaction  constant
          $j=0.01$  and  different  values  of  the  anisotropy  $q$.
          The diagram for $q = 0$ is representative of a finite but
          infinitely small anisotropy (see the text). The dashed lines are
          the {\it disorder line} as explained in the text. 
          Inserts show the order 
          parameters for the selected $x$ values (vertical dashed lines).}}
\end{figure}

\begin{references}
\bibitem{Tranquada88} J.M. Tranquada, {\it et al.}
Phys.Rev.Lett. {\bf 60}, 156 (1988).
\bibitem{Kadowaki88} H. Kadowaki, {\it et al.}
Phys. Rev. B {\bf 37}, 7932 (1988).
\bibitem{Shamoto93} S. Shamoto, {\it et al.}
Phys. Rev. B {\bf 48}, 13817 (1993).
\bibitem{Moudden88} A.H. Moudden, {\it et al.}
Phys. Rev. B {\bf 38}, 8720 (1988).
\bibitem{Li90} W.-H. Li, {\it et al.}
Phys. Rev. B {\bf 41}, 4098 (1990).  
\bibitem{Casalta94} H. Casalta, {\it et al.}
Phys. Rev. B {\bf 50}, 9688 (1994).
\bibitem{Brecht95} E. Brecht, {\it et al.}
Phys. Rev. B {\bf 52}, 9601 (1995).
\bibitem{Brecht97} E. Brecht, {\it et al.} (unpublished).
\bibitem{Miceli89} P.F. Miceli, {\it et al.} 
Phys. Rev. B {\bf 39},12375 (1989).
\bibitem{Mirebeau94} I. Mirebeau, {\it et al.}
Phys. Rev. B {\bf 50}, 3230 (1994).
\bibitem{Schmidt97} O. Schmidt {\it et al.} (unpublished).  
\bibitem{Rossat-Mignod93} J. Rossat-Mignod, {\it et al.}
Physica B {\bf 192}, 109 (1993).
\bibitem{Hayden96} S. Hayden, {\it et al.}
Phys. Rev. B {\bf 54}, 6905 (1996)
\bibitem{Brecht96} E. Brecht, {\it et al.}
Physica C {\bf 265}, 53 (1996).
\end{references}
\end{document}